\documentclass[conference]{IEEEtran}
\pdfoutput=1
\IEEEoverridecommandlockouts
\usepackage{cite}
\usepackage{amsmath,amssymb,amsfonts}
\usepackage{algpseudocode}
\usepackage{graphicx}
\usepackage{booktabs}
\usepackage{textcomp}
\usepackage{xcolor}

\usepackage[ruled,vlined]{algorithm2e}
\usepackage[compact]{titlesec} 
\begin{document}

\title{Blind Image Denoising and Inpainting Using Robust Hadamard Autoencoders\\}

\author{\IEEEauthorblockN{Rasika Karkare\IEEEauthorrefmark{1}\\}
\IEEEauthorblockA{Data Science\\
Worcester Polytechnic Institute\\
Worcester, MA 01609\\
Email: rskarkare@wpi.edu}
\and
\IEEEauthorblockN{Randy Paffenroth\IEEEauthorrefmark{2}\\}
\IEEEauthorblockA{Mathematical Sciences, \\
Comp. Science and Data Science\\
Worcester Polytechnic Institute\\
Worcester, MA 01609\\
Email: rcpaffenroth@wpi.edu}
\and
\IEEEauthorblockN{Gunjan Mahindre\IEEEauthorrefmark{3}\\}
\IEEEauthorblockA{Electrical and Computer Engineering\\
Colorado State University\\
Fort Collins, CO 80523\\
Email: gunjan.mahindre@colostate.edu}
}


\maketitle

\begin{abstract}
In this paper, we demonstrate how deep autoencoders can be generalized to the case of inpainting and denoising, even when no clean training data is available.  In particular, we show how neural networks can be trained to perform all of these tasks simultaneously. While, deep autoencoders implemented by way of neural networks have demonstrated potential for denoising and anomaly detection, standard autoencoders have the drawback that they require access to clean data for training. However, recent work in Robust Deep Autoencoders (RDAEs) shows how autoencoders can be trained to eliminate outliers and noise in a dataset without access to any clean training data.  Inspired by this work, we extend RDAEs to the case where data are not only noisy and have outliers, but also only partially observed.  Moreover, the dataset we train the neural network on has the properties that all entries have noise, some entries are corrupted by large mistakes, and many entries are not even known. Given such an algorithm, many standard tasks, such as denoising, image inpainting, and unobserved entry imputation can all be accomplished simultaneously within the same framework. Herein we demonstrate these techniques on standard machine learning tasks, such as image inpainting and denoising for the MNIST and CIFAR10 datasets.  However, these approaches are not only applicable to image processing problems, but also have wide ranging impacts on datasets arising from real-world problems, such as manufacturing and network processing, where noisy, partially observed data naturally arise.
\end{abstract}

\begin{IEEEkeywords}
\\Autoencoders, Robust Deep Autoencoders, Blind Denoising, Inpainting, Anomaly Detection\\
\end{IEEEkeywords}

\section{Introduction}
Recent advances in deep learning and neural networks have shown the effectiveness of these techniques in numerous fields such as object detection, image recognition, drug discovery, and genomics to name but a few~\cite{5}.
By using backpropogation, deep learning demonstrates how the parameters of the network can be changed to compute the representations in each layer. Such models have proven to learn faithful representations of a dataset by learning non-linear features in the data.  
Multiple layers are used to learn different levels of abstraction in the data and are essentially what make the network \emph{deep}~\cite{lecun2015deep}. 

For image reconstruction, standard deep autoencoder variations such as denoising autoencoders are widely used~\cite{2}. However, the drawback of using such models is that they require access to clean data for training which is not readily available in all real-world problems. In this work we extend such techniques to the task of image denoising and inpainting where we do not have access to the  \emph{noise-free} version of the data. RDAEs~\cite{4}, as mentioned previously, do not require clean data for training overall, however, these models isolate the noise and outliers in the input and the autoencoder is trained after this isolation. 

Moreover, an important issue with RDAEs~\cite{4}, in addition to the fact that standard RDAEs cannot be applied when the data is only partially observed,  is that while the overall method does not require clean training data, the neural network that has been embedded in the larger method does require clean training data.  The structure of the RDAE equations that leads to this issue is perhaps not readily apparent, and we will discuss it in detail in the background section below.  While not an issue if data is always processed in a single batch, this does require the retraining of the entire neural network if the data should ever change. The ability to denoise new data that the network was not originally trained on, what we herein call \emph{inference}, is not possible using the RDAEs formalism~\cite{4}. Accordingly, herein we extend the framework of the RDAEs model in a way such that it can be tested on unseen data that it was not originally trained on.

A main focus of this paper is an attempt to overcome the above drawback in RDAEs so that they can be used in a more general context.  As we explain in the further sections of the paper, our model is trained only with the corrupt data as input to the neural network, as opposed to RDAEs~\cite{4}. This makes our model \emph{inductive}, in which it can \emph{infer} the noise in the data, as opposed to RDAEs. Our model, which we call a Robust Hadamard Autoencoder (RHA) can also handle data that is only partially observed, which is a feature that is lacking in standard RDAEs. It is interesting to note that our RHA is the first method, of which we are aware, that provides all of the functionality of a classic linear Robust Principal Component Analysis (RPCA) type algorithm~\cite{6,13}, but in the context of a non-linear autoencoder that can simultaneously inpaint and denoise data in a \emph{blind} manner. All the code that generates the results from this paper is freely available at github \footnote[1]{https://github.com/rskarkare/Robust-Hadamard-Autoencoders}.

\section{Contribution} 
In this work, we aim to generalize RDAEs to a case where we make the model \emph{testable} by  providing only corrupt data as input to the neural network. We demonstrate the superior performance of our method as compared to the state-of-the-art techniques, using the standard MNIST and CIFAR10 datasets in presence of noise as well as missing regions in the data. Moreover, our model is able to \emph{infer} the noise in the data and successfully eliminate it, without being given any information about the \emph{noise-free} version of the data. As we demonstrate in this paper, our model is robust and can be used in presence of random impulse noise as well as coherent corruptions such as inferring random blocks of missing data.

Our model differs from existing techniques as follows:

\begin{enumerate}
\item Our model is able to obtain a \emph {non-linear} projection of the data, unlike RPCA which can only obtain a \emph {linear} projection to a lower dimension.
\item	We do not need clean training data to train our model unlike standard techniques such as denoising autoencoders~\cite{2}. 
\item	We do not provide our model with low-dimensional input, as is the case with RDAEs. The primary difference between the two models being, that our neural network model \emph{infers} the noise and successfully eliminates it, whereas, in the case of RDAEs, the neural network model is given data in which the noise has been subtracted off during the training process. 
\item We also address the issue of filling-in missing blocks of data based on the surrounding information~\cite{28,12}. We demonstrate the superior performance of our model on denoising as well as image inpainting, as compared to the RDAEs model, which cannot handle coherent corruptions in the data, such as missing blocks of data in images. We also compare our model to a state-of-the-art inpainting algorithm, namely, the context encoder (CE)~\cite{9}, and show that our model shows comparable performance with the CE model for the task of inpainting in presence of missing data and outperforms CE in the case of denoising and inpainting simultaneously. This is despite the fact that in the CE model, the network is given the clean block of data as a reference for training the generator in the Generative Adversarial Network (GAN), whereas, our model does not have any information about the \emph{missing part} in the data (i.e. the complete version of the image). 
\item We compare our model with a standard autoencoder (SAE) and a variational autoencoder (VAE)~\cite{vae} for the task of image denoising and inpainting on the CIFAR and MNIST datasets. Our model outperforms both of the other models in this task, as will be demonstrated in the results section. 
\end{enumerate}

\section{Background}
Here we cover some background information about Deep autoencoders and RPCA, which are the foundations of our model. 
\subsection{Autoencoders:}
An autoencoder is a type of neural network that is primarily used to learn an identity map for the input data. Autoencoders can be used to obtain a compressed representation of the data by projecting it to a low-dimensional latent layer~\cite{5}.
Along with learning a low-dimensional representation, deep autoencoders are made non-trivial due to the non-linearities that are introduced in the hidden layer by using various activation functions. The network learns a map of the input through encoding and decoding network pairs and the reconstruction is given by:

 \begin{equation}
     \bar{X} = D(E(X)).
 \end{equation}

Furthermore, the autoencoders give us a solution to the below optimization problem.

\begin{equation}
   \min_{D,E} ||X-D(E(X))||_F^2 \hspace{0.05cm}.
\end{equation}
\noindent Here, $E$ is the encoded representation of the input data to the hidden layer and $D$ is the decoded representation from the hidden layer to the output layer. $\bar{X}$ is the reconstructed version of the input data. The goal is to minimize the difference between the original input data and the reconstructed data. The loss function is commonly chosen to be the Frobenius norm between the original and reconstructed data. In order to learn complicated distributions, deep learning models use multiple hidden layers to learn higher order features in the data~\cite{5}. Unfortunately, the presence of noise and outliers in the data affect the quality of the features that are learnt by the autoencoder~\cite{26,27}.

\subsection{Robust Principal Component Analysis (RPCA):}

Principal Component Analysis (PCA) is a linear dimensionality reduction technique that is widely used in numerous applications. However, a drawback of PCA is that it is grossly sensitive to outliers and noise and hence a variation, namely Robust Principal Component Analysis (RPCA), is used in the presence of outliers and noise. RPCA allows separation of the outliers and anomalies in such a way that the \emph{noise-free} data can be recovered~\cite{6}.
RPCA splits the input data $X$ as $L~+~S$, where $L$ is the low-rank matrix and $S$ is the sparse matrix. Here $X,L,S \in~\mathbb{R}^{m \times n}$.\par
$L$ contains the low-dimensional representation of $X$ and $S$ contains the element-wise outliers in the data. The optimization problem can be written as follows:\\
\begin{equation}
    \begin{split}
     &\min_{L,S} \rho (L) + \lambda ||S||_0 \\
     &\mathrm{s.t.}~||X-L-S||^2_{F} = 0.
    \end{split}
\end{equation}
\noindent Here $\mathrm{\rho}(L)$ is the rank of $L$ and $||S||_0$ is the number of non-zero entries in $S$, $||F||$ is the frobenius norm. $||S||_1$ is the $l_1$-norm of $S$~\cite{24}.

However, since the above optimization is NP-hard, the relaxed version of the class of problems is given by the $||.||_1$ norm~\cite{6}.

The convex version of the problem is thus given as follows:\\
\begin{equation}
\begin{split}
    &\min_{L,S} ||L||_{\ast} + \lambda ||S||_1 \\
    &\mathrm{s.t.}~||X-L-S||^2_{F} = 0.
\end{split}
\end{equation}\\

\noindent Thus, RPCA allows one to learn a more faithful representation of the \emph{noise-free} low-dimensional representation of the data by carefully separating out the sparse outliers~\cite{13}. 
In our model, the nuclear norm of RPCA is replaced by the reconstruction error of an autoencoder. This gives us a non-linear projection to a low-dimensional hidden layer as opposed to a linear projection given by RPCA.  

\subsection{Robust Deep Autoencoders (RDAEs):}

Robust Deep Autoencoders combine the salient features of autoencoders and RPCA, to solve the following optimization problem:

\begin{equation}
\begin{split}
    \min_{\theta} ||L_D - D_{\theta}(E_{\theta}(L_D)))||_2 &+ \lambda ||S||_0 \\
    \mathrm{s.t.}~X-L_D-S &= 0.
    \end{split}
\end{equation}

However, since the above term is not computationally tractable, the $l_{0}$-norm is replaced with the $l_{1}$-norm and the convex relaxation of this problem similar to RPCA is given below~\cite{4}:
\begin{equation}\label{rdae}
\begin{split}
    \min_{\theta} ||L_D - D_{\theta}(E_{\theta}(L_D)))||_2 &+ \lambda ||S||_1 \\
    \mathrm{s.t.}~X-L_D-S &= 0.
\end{split}
\end{equation}
\emph{As it can be seen in the above optimization, the RDAEs model require the input to the neural network model to be low-rank$(L_D)$}. Moreover, it has a limitation that it is not inductive and cannot be used to filter out noise and anomalies on unseen data without having to retrain the network on the new data. In this work, we aim to generalize the RDAEs by developing a model that is robust to noise and anomalies, while at the same time it is inductive and can be tested on unseen corrupt images, because the only input to the neural network is a corrupt image as opposed to a low-rank one in the case of RDAEs. Our model is inspired by RPCA and RDAEs wherein, the input data X is comprised of two parts $X = L + S$ where $L$ can be effectively reconstructed by the autoencoder and $S$ contains the outliers and noise in the data.  Our model also addresses the issue of filling-in missing blocks of data by using a Hadamard or element-wise product in the cost function~\cite{28}.  This has important applications in fields such as missing value imputation in manufacturing datasets. Moreover, in such datasets some instances do not have the complete set of features available during training. Rather than eliminate the rows that lack this information, we want to be able to reproduce the values of the missing features that are as close to the original as possible. Typically, the values of certain parameters found in such datasets are missing at random. Hence, we demonstrate the results of our model on standard benchmark image datasets, namely the MNIST and CIFAR10. It can be seen that our model performs better as compared to the state-of-the-art techniques on these datasets in case of denoising and inpainting simultaneously.

\section{Methodology}
In this section we explain our objective function, for both the denoising as well as the task of image inpainting. The key idea as inspired by the above models is that the autoencoder would generate a low-dimensional non-linear representation of the \emph{low-rank} input and filter out the noise, which is incompressible. The noise and anomalies will essentially be filtered out in the sparse matrix as per the analogy with RPCA. Thus, we combine the non-linearity capability of an autoencoder with the outlier detection capability of RPCA to obtain an objective function for our model. The key difference being that the input to the neural network which is embedded inside the entire network as a whole, is a corrupt image. This gives rise to \emph{blind} denoising capabilities of our model, wherein no information about the noise in the data is explicitly provided to the neural network. 

\begin{figure*}
    \centering
    \includegraphics[width=0.88\textwidth]{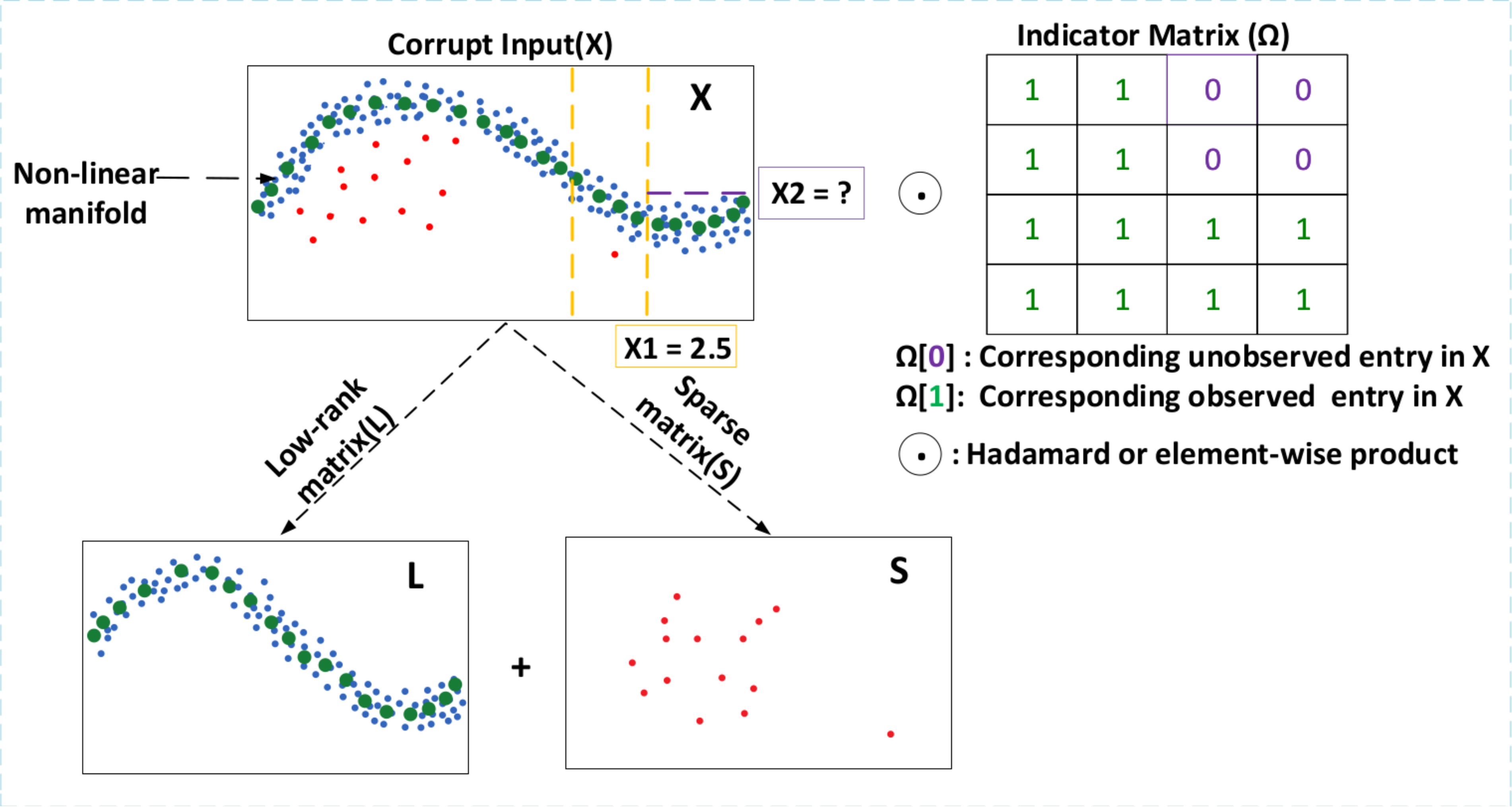}\\
    \caption{Our model combines the non-linearity capability of an autoencoder with the denoising capability of RPCA. In addition to random impulse noise, it is also robust to coherent corruptions in the data such as blocks of missing values which we impute using the Hadamard or element-wise product in our objective function. Every entry in the dataset that is unobserved, will have a corresponding entry of 0 in the indicator matrix and every entry that is observed will have a corresponding entry of 1 in the indicator matrix. Thus, we mask the input by multiplying element-wise with the indicator matrix. We then replace all the missing entries with the mean value of the observed data before providing the corrupt input to the autoencoder network. Note that the mean values are purely for initialization and are updated to the imputed values by the neural network.}
    \label{fig:plot_1}
    \vspace{-0.3cm}
\end{figure*}
\subsection{Robust Hadamard Autoencoders with $l_1$ regularization:}
Figure~\ref{fig:plot_1} shows the developmental ideas of our model.  
Inspired by RPCA,  we want to obtain a low-dimensional representation of our data without the few data points that are considered to be the \emph{outliers or noise}. These points are filtered out in the sparse matrix as they are incompressible. The nuclear norm in the RPCA objective function is replaced by the reconstruction error of an autoencoder in order to obtain a non-linear representation of the input. 
Our objective function thus becomes, 

\begin{equation}\label{rha}
    \min_{\theta} ||(X-S)-D(E(X)))\odot \Omega||_2  + \lambda ||S||_1.
\end{equation}
Here, similar to RPCA, we replace the $l_{0}$ norm on $S$ with the $l_{1}$ norm, in order to make it computationally tractable. 
The $\odot$ represents the Hadamard or element-wise product between the cost function and $\Omega$~\cite{28}, $\Omega$ being an indicator matrix of ones and zeros. Every entry in the dataset that is missing has a corresponding entry of zero in the indicator matrix, whereas every entry in the dataset that is observed has a corresponding entry of one in the indicator matrix. 
The idea is that the cost function is calculated with respect to only the data that is observed. For data that is missing, the corresponding entries in the indicator matrix are zero and thus the missing entries are ignored by the cost function. The assumption here is that, while imputing the data if the autoencoder does well on the data that is observed, then the autoencoder will also do well on the data that is missing. The $\lambda$ parameter in the objective function similar to RDAEs needs to be tuned in order to change the penalty on the sparse matrix to filter out the noise. As seen in Eq.~\ref{rha}, the input to the neural network in this case is the $X$ matrix, which is the corrupt version of the data. The target or reference clean image is not provided, as is the case with a standard denoising autoencoder. Moreover, the autoencoder learns the $L$ and  $S$ matrix by \emph{inferring} the noise during training, and it is then filtered out from the input to obtain a low-rank representation of the data (i.e. $L~=~X-S$). Note here that in Eq.~\ref{rdae} for the RDAEs model the input to the neural network is the low-rank matrix given by $L_D$ in the objective function. This illustrates the difference between the two models.
For the task of pure image inpainting, the above objective function can be easily modified to the one below:

\begin{equation}
\min_{\theta} ||(X-D(E(X)))\odot \Omega||_2.
\end{equation}

Here the $S$ term is not needed because there is no noise to be filtered out in this case. 
This objective function ignores the regions in the data that are missing during optimization and only finds the optimal values for imputation based on the regions that are observed during training.
\section{Algorithm Training}

For training our model we performed hyper-parameter tuning to find the optimal set of parameters for our model. This section details the parameters that we used to build our model as well as the training method that we used to optimize the network. We use an autoencoder network with two hidden layers. The number of nodes in the hidden layers is 200 and 50 respectively and the batch size is 40. It can be seen in Fig.2 that our model is fairly robust across multiple batch sizes.  We use the sigmoid  activation function owing to the fact that the data was normalized between 0 and 1 during the pre-processing phase. It is interesting to note that even such a simple network, when armed with additional terms from RPCA~\cite{6, 7, 28} can lead to superior performance over much deeper networks. Of course, studying how such approaches work with even deeper networks is a direction for future research. The next section details the method used for corruption and training of the algorithm.\\

For corrupting the image, we add salt and pepper noise to each of the images in the input and also mask random blocks by replacing the pixels in these blocks by the mean pixel value, i.e., 0.5. The mean initialized values are then imputed using our model. Salt and pepper noise, also known as impulse noise is a form of sparsely occurring noise in an image signal where randomly selected pixels are black and white~\cite{22, 25}. The learning rate used for all our experiments is 0.01 and the optimizer is Adam. The pre-processed corrupt data is then fed as input to the neural network. We train for 50 epochs and then test on a different set of corrupt images which we call our \emph{test dataset}. We use an optimization similar to that used by RDAEs wherein we use a combination of backpropagation and proximal gradients as suggested in~\cite{4}.The $l_{1}$ norm can be optimized efficiently through the use of a proximal method as given in ~\cite{4}.
Below is the pseudo-code of the algorithm that is used to train our model. For additional details refer ~\cite{7}.
\begin{algorithm}
\SetAlgoLined
\caption{Proposed Training Method}
Input: $X \in~\mathbb{R}^{m \times n}$. Here $X$ is the corrupt image. We define $L_S$ = $X$ as the initial state which will be updated during training.\\
Initialize $L_D \in~\mathbb{R}^{m \times n}$, $S \in~\mathbb{R}^{m \times n}$ to be zero matrices. Here $L_D$ is the low-rank matrix and $S$ is the sparse matrix.
\While{true}{
\begin{itemize}
  \item Provide the autoencoder with $X$ as the input for training.\\
    \item Minimize $||(X-S)-D(E(X))||_2$ using backpropagation.\\
    \item Set $L_D$ to be the reconstruction from the trained \indent autoencoder: $L_D~=~D(E(X ))$
    \item Set S to be the difference between $X$ and $L_D$: $S = X-L_D$\\
    \item Optimize $S$ using a proximal operator: $S~=~prox_{\lambda, l_{2,1}(S)}$ or  
    \indent $S~=~prox_{\lambda,l_{1}}(S)$\\
    \item Check the convergence condition that $L_D$ and $S$ \\ are \indent close to the input $X$ thereby satisfying the \\ constraint: \indent $c_1~=~ ||X-L_D-S||_2 /||X ||_2$\\
    \item Check the convergence condition that $L_D$ and $S$ \\ have converged to a fixed point: $c_2~=~||L_S-L_D-S||_2 / ||X||_2$\\
    if $c_1 < \epsilon~or~c_2 < \epsilon$:\\
    ~~~~~~break\\
    \item Update $L_S$ for convergence checking in the next iteration:
    \indent $L_S~=~L_D+S$
\end{itemize}
    }
Return $L_D$ and $S$
\end{algorithm} 
\section{Results}

Table~\ref{tab:table1} shows a comparison between all the models mentioned above including SAE, RDAEs, VAEs and CE with our model for the task of image denoising and inpainting on the CIFAR10 dataset. 
\begin{table}[ht!]
\centering
\caption{Comparison of all models on the task of simultaneous image inpainting and denoising on the CIFAR10 dataset} 
\begin{tabular}{c c c c c c}
\hline
\textit{Corruption} & \textit{SAE} & \textit{VAE} & \textit{RDAE} & \textit{CE} & \textit{RHA}\\
\hline
10 & 0.02992 & 0.06718  & 0.01895 & 0.01763 & \textbf{0.01601}\\
70 & 0.05672 & 0.10521  & 0.04795 & 0.06176 & \textbf{0.03917}\\
350 & 0.27894 & 0.19901 & 0.15712 & 0.45005 & \textbf{0.13724}\\
\hline
\end{tabular}
\label{tab:table1}
\vspace{-0.3cm}
\end{table}\\

We have used three values of corruption to demonstrate the superior performance of our model as compared to any of the other models on the task of denoising and inpainting simultaneously. The corruption column indicates the number of pixels that are corrupted by the addition of random salt and pepper noise. Along with this we also add random 10x10 blocks of missing regions to all the input images. The choice of corruption shown in the table is based on using the smallest, largest and intermediate values of corruption, to show that our model outperforms the other methods across a wide range of corruption values. For the methods that have a $\lambda$ parameter in their objective function, we use the $\lambda$ that gives the best performance, i.e., the $\lambda$ value with the smallest RMSE as the final metric reported in the table.
\begin{table}[ht!]
\centering
\caption{Comparison of all models on the task of simultaneous denoising and inpainting with different sizes of missing data on the CIFAR10 dataset.}
\resizebox{0.495\textwidth}{!}
{
\begin{tabular}{c c c c c c c}
\hline
\textit{Corruption} & \textit{Size} & \textit{SAE} & \textit{VAE} & \textit{RDAE} & \textit{CE} & \textit{RHA}\\
\hline
70&5x5 & 0.03887 & 0.06183 & 0.03896 & 0.04523 & \textbf{0.03461}\\
70&10x10 & 0.05672 & 0.10521 &  0.04795  & 0.06176 & \textbf{0.03917}  \\
70&20x20 & 0.23886 &  0.37617  &  0.19521 &  0.17782 & \textbf{0.14361}  \\
\hline
\end{tabular}
}
\label{tab:table2}
\vspace{-0.3cm}
\end{table}\\

Table ~\ref{tab:table2} summarizes a comparison of all the models on the task of denoising and inpainting simultaneously on the CIFAR10 dataset. Here, we fix the level of corruption and test how well our model compares to the state-of-the-art techniques on different sized blocks of missing data. Corruption indicates the level of corruption in terms of addition of salt and pepper noise. The size column indicates the block sizes of missing data. We can see that across multiple sizes of missing data, our model shows superior performance as compared to any of the other techniques in terms of the reconstruction error between the original image and recovered image using the respective models.
Now we demonstrate the results of our model on both the MNIST and the CIFAR10 datasets~\cite{17,18}. For the grayscale MNIST dataset, each image has a shape of 28x28, whereas for the CIFAR10 colored dataset each image has a shape of 32x32x3 owing to the three color channels. The metric used for comparison in each of the results shown is the percentage difference in Root Mean Squared Error (RMSE) between the two models. It is given by the formula:
\begin{equation}
\begin{split}
&Percentage(\%)~difference~in~RMSE = \\ 
&\frac{RMSE(model A) - RMSE(model B)}{RMSE(model A)} \times 100.
\end{split}
\end{equation}

Here model A refers to the model that we are comparing against and model B refers to our model. Thus, higher the difference between the two, the better our model performs. 
In the colormaps shown, the corruption level indicates the number of pixels that are corrupted with the salt and pepper noise in each image. For example, in the MNIST dataset a corruption level of 10 indicates that 10 out of 784 pixels are corrupt in every image. In addition, for inpainting and denoising simultaneously we use random blocks of size 10x10 pixels and replace them with the mean value, i.e., 0.5, in addition to the salt and pepper corruption before feeding the corrupt images as input to the neural network. In order to have a fair comparison among the methods, we use the same amount of corruption and maintain a constant size of the missing blocks, i.e., 10x10 in all the models. To evaluate how RHA performs under different sizes of missing data we also tested it under the conditions of different sizes of random missing blocks. The results for this are given in Table ~\ref{tab:table2}.
\subsection{Comparison with RDAEs for denoising}
\begin{figure}
    \centering
    \includegraphics[width=0.475\textwidth]{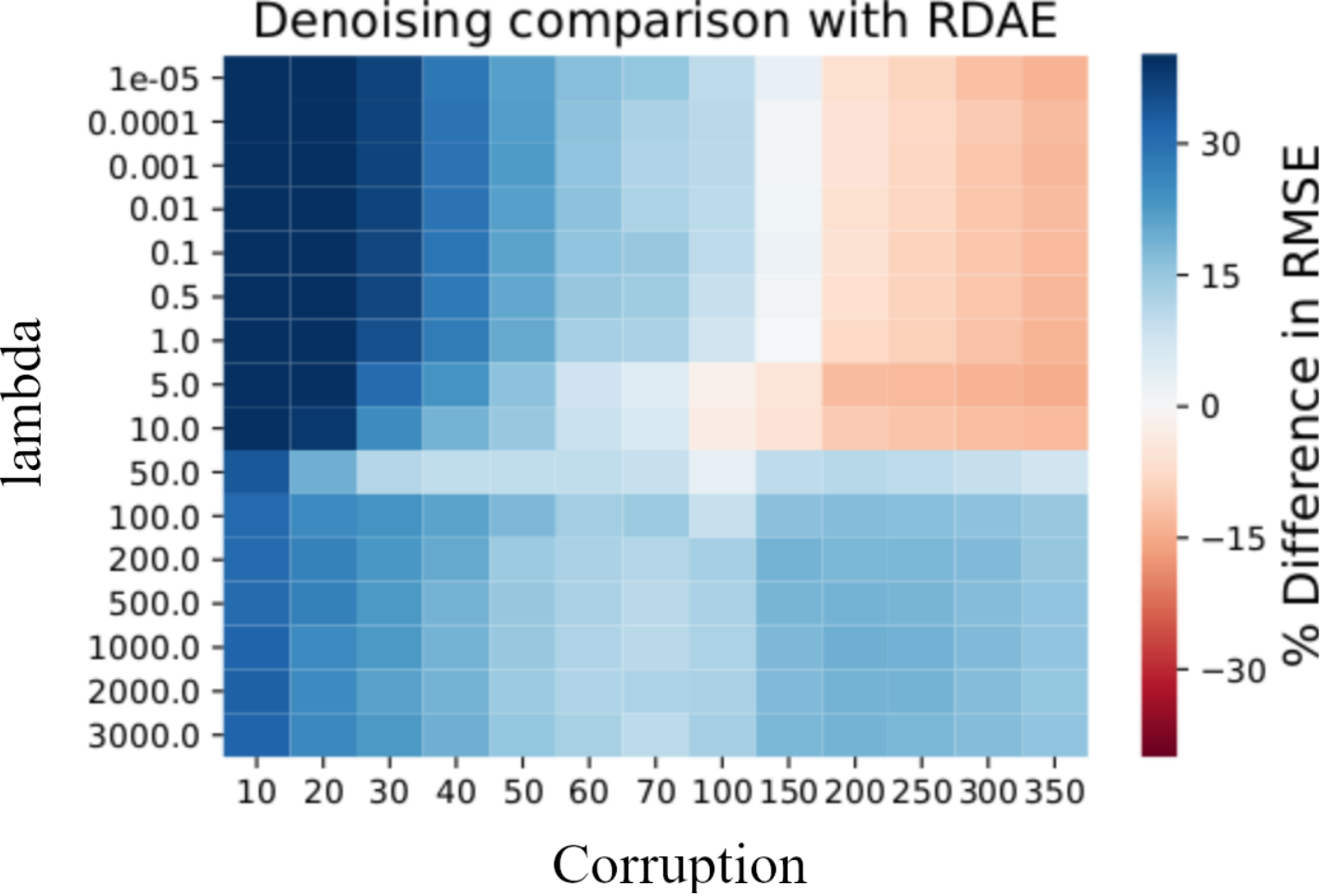}
    \caption{Comparison of RHA with RDAEs for the task of denoising on the MNIST dataset. The metric used for comparison in this case is Root Mean Squared Error (RMSE) between the reconstructed image and the  \emph{noise-free} image. We calculate the percentage (\%) differences between the RDAEs~\cite{4} model and RHA for different values of corruption and $\lambda$ on the test dataset. Thus, higher the difference between the two models, the better our model performs. \emph{It can be seen that with the right selection of the parameter $\lambda$, our model is able to outperform the RDAEs model across all levels of corruption.}}
    \label{fig:plot_2}
    \vspace{-0.3cm}
\end{figure}
Here we compare the denoising capabilities of our model with that of RDAEs. 
Figure~\ref{fig:plot_2} shows a comparison of our model with the RDAEs model for the task of denoising on the MNIST test dataset.  It can be seen that with the right selection of parameter $\lambda$, our model is able to outperform the RDAEs model across all levels of corruption. The code used  for the RDAEs can be obtained at github \footnote[2]{https://github.com/zc8340311/RobustAutoencoder}.
\subsection{Comparison with Context Encoder (CE) for inpainting}
\begin{figure*}[hbt!]
    \centering
    \includegraphics[width=\textwidth]{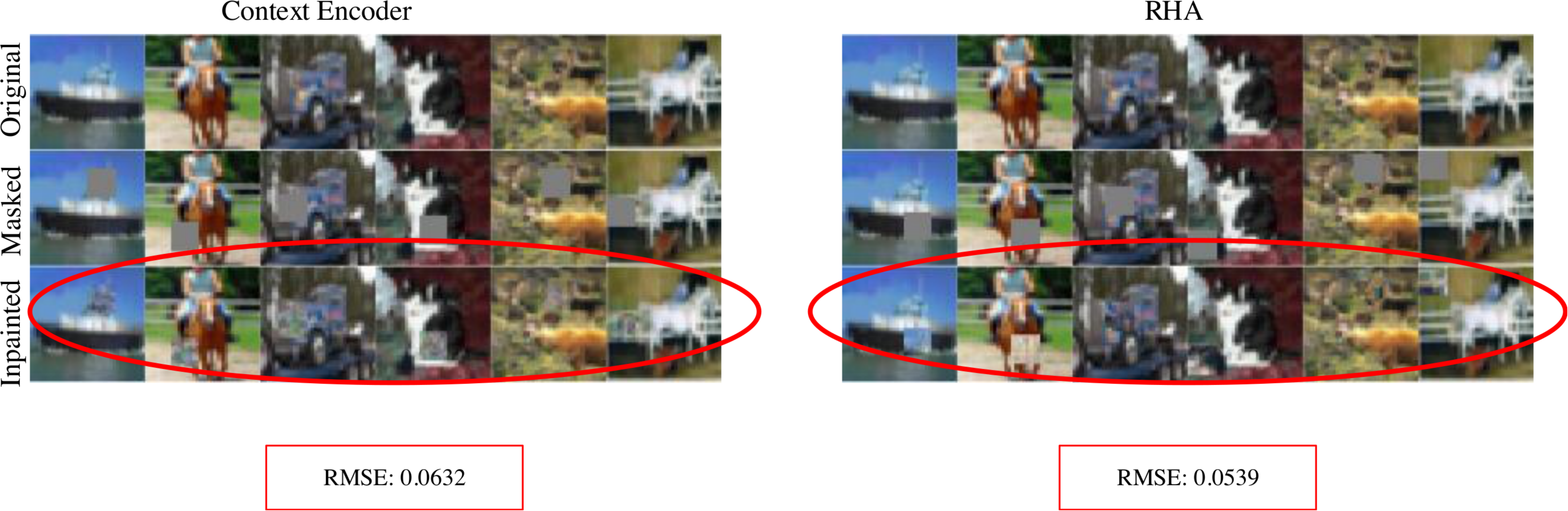}
    \caption{Comparison of our model with Context Encoder (CE)~\cite{conae} for image inpainting on the CIFAR10 dataset. This comparison uses RMSE as a metric between the inpainted image and the original image. It can be seen that our model outperforms the CE model in the task of image inpainting. It is important to note that in case of the CE model, the model uses the original image as a reference for training the generator in the GAN network, whereas our model does not use the original image as reference and fills in the missing regions only based on the surrounding pixel information.}
    \label{fig:plot_3}
    \vspace{-0.3cm}
\end{figure*}
Figure~\ref{fig:plot_3} shows a comparison of our model with that of a state-of-the-art inpainting algorithm, namely, the Context Encoder (CE) which essentially uses a GAN model for inpainting missing blocks in the CIFAR10 dataset. It can be seen from the comparison of the models showing the inpainted images in the third row that our model outperforms the CE algorithm despite not having any information about the missing regions in the data, as opposed to the CE model. In the CE framework, the missing regions are regressed onto the \emph{ground-truth} version of the image during training. 
\subsection{Comparison with a Standard Autoencoder (SAE)}
\begin{figure}[hbt!]
    \centering
    \includegraphics[width=0.475\textwidth]{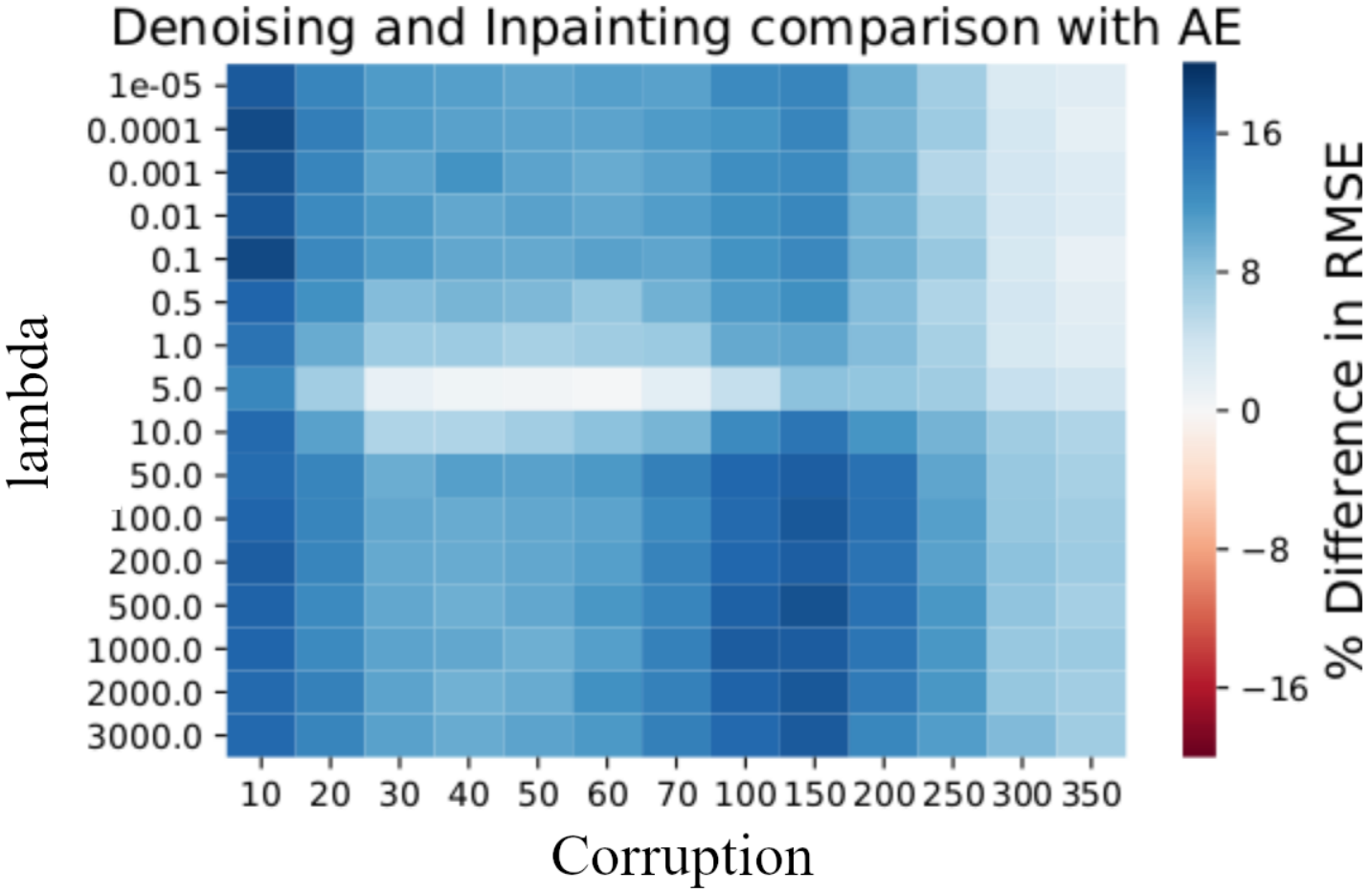}
    \caption{Comparison of our model with a SAE in presence of noise and missing blocks of data in the MNIST dataset. The metric used for comparison is the percentage (\%) difference in RMSE between the reconstructed image and the original \emph{noise-free} image. It can be seen from the colormap that we outperform SAE across all corruption levels shown.}
    \label{fig:plot_4b}
    \vspace{-0.3cm}
\end{figure}
In order to assess the denoising and inpainting capabilities of a standard autoencoder owing to its ability to go low-dimensional in the hidden layers, we compare the performance of a SAE with our model. Figure~\ref{fig:plot_4b} shows a comparison of our model with a SAE across multiple values of $\lambda$ and corruption in terms of the RMSE on the MNIST dataset. It can be seen that our model outperforms a SAE in terms of the reconstruction error across all the levels of corruption as seen on the colormap. The SAE, although faithful to the original input, incorrectly reproduces the noisy data. 
\subsection{Comparison with RDAEs for the task of denoising and inpainting simultaneously}
\begin{figure}[hbt!]
    \centering
    \includegraphics[width=0.475\textwidth]{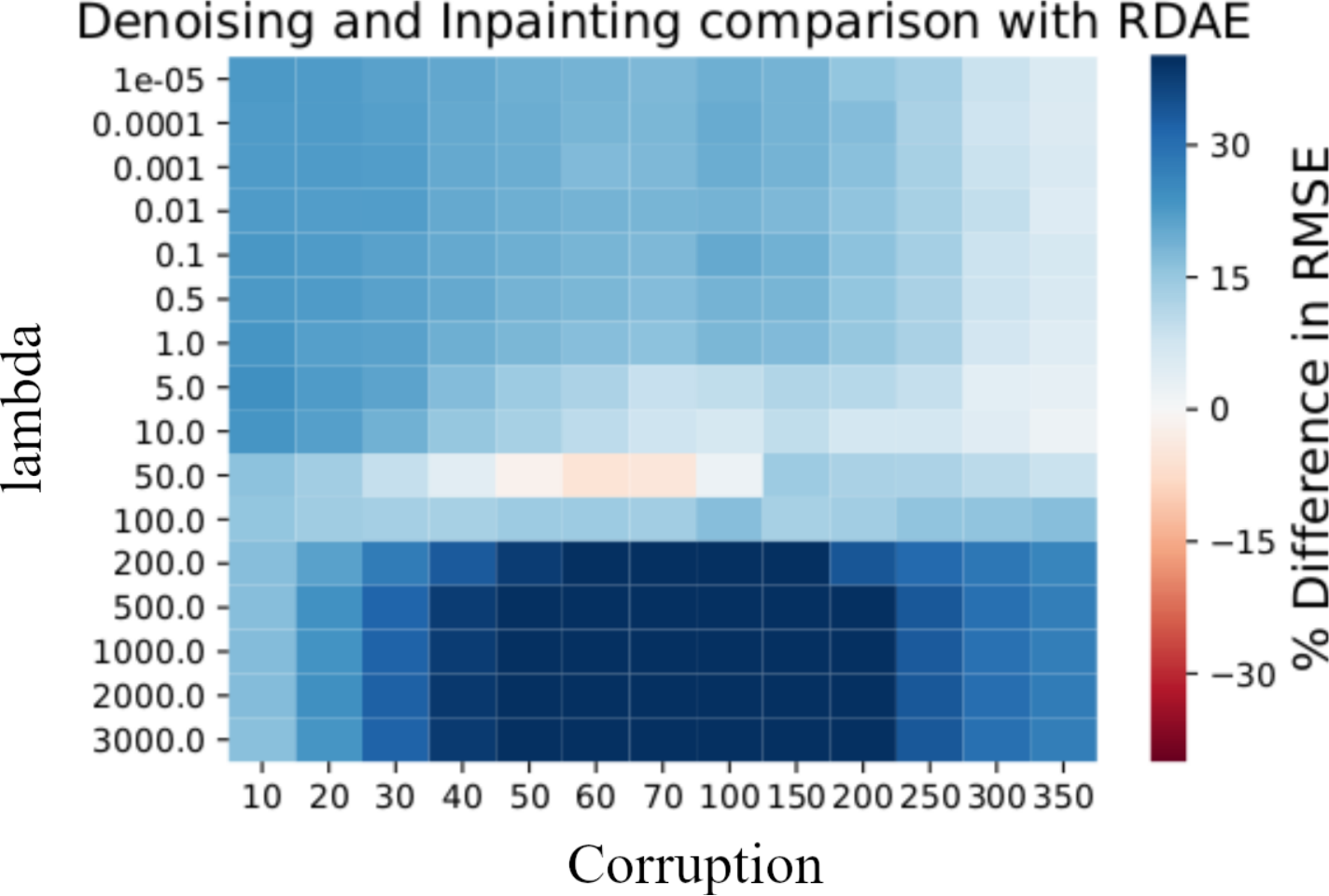}
    \caption{Comparison of our model with the RDAEs model for the task of denoising and inpainting simultaneously on the MNIST dataset. The metric used for comparison in this case is RMSE between the original \emph{noise-free} image and the reconstructed image using the respective models. The metric is based on percentage (\%) differences in the RMSE between the RDAEs model and our model. Thus, higher the difference between the two, as indicated by the positive values on the colormap, the better our model performs. The results on the test dataset show that we outperform the RDAEs in the task of denoising and inpainting simultaneously for all levels of corruption in the testing data by picking the right value of the parameter $\lambda$.}
    \label{fig:plot_5}
    \vspace{-0.3cm}
\end{figure}
Figure~\ref{fig:plot_5} shows a comparison of our model with the RDAEs model in the presence of noise and missing data on the MNIST dataset. It can be seen that we outperform the RDAEs model in terms of RMSE across all levels of corruption. Here, higher the difference, the better our model performs as compared to the RDAEs model. 
\subsection{Comparison with CE model for the task of denoising and inpainting simultaneously}
\begin{figure}[!hbt]
    \centering
    \includegraphics[width=0.495\textwidth]{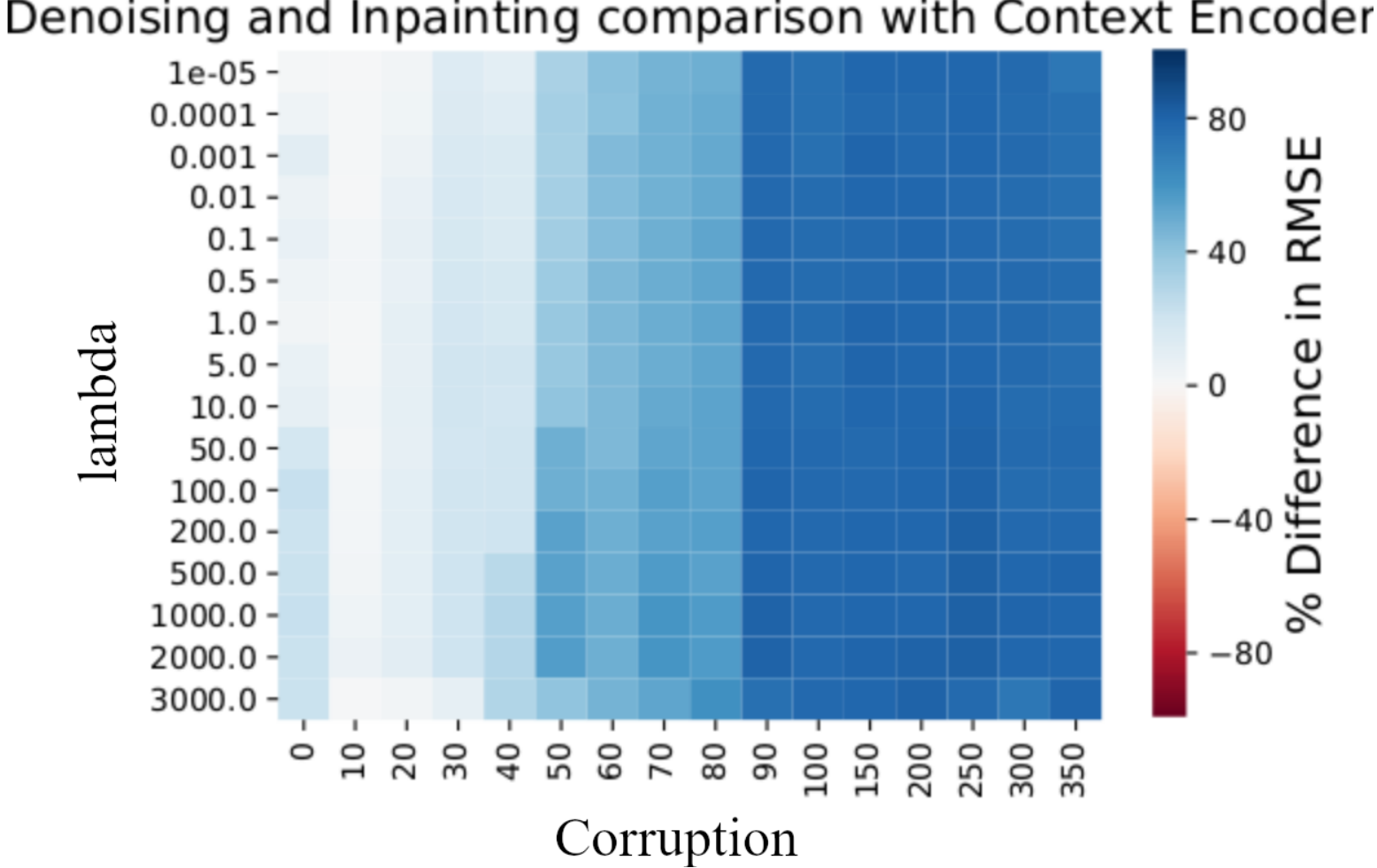}
    \caption{Comparison of our model with the CE in presence of noise and missing blocks of data on the CIFAR10 dataset. It can be seen that despite using a clean \emph{noise-free} version of the image as reference for training the generator in the GAN network, the CE model fails to reconstruct the clean image when noise is added to the input. Our model shows superior performance and outperforms the CE model when the same amount of noise is added to both models. The metric used here for comparison is the RMSE between the original \emph{noise-free} image and the reconstructed image using the respective models. It can be seen from the colormap above that although the differences are not significant at lower levels of corruption, our model notably outperforms the CE model as the corruption level increases. The colormap indicates the percentage differences between the RMSE for the CE model and our model, thus higher the differences, the better our model performs as compared to the CE model. The architecture used to train the CE model is as given in ~\cite{conae}.}
    \label{fig:plot_6}
    \vspace{-0.3cm}
\end{figure}
Figure~\ref{fig:plot_6} shows a comparison between our model and a state-of-the-art inpainting algorithm, namely the CE model in terms of RMSE on the CIFAR10 dataset. It can be seen that although the CE model shows almost comparable performance with ours for the task of inpainting alone, the network does not perform well with the addition of noise and the generator fails to reconstruct the clean images. The architecture of the CE network is the same as given in ~\cite{9, conae} with the change that we add random salt and pepper noise to the input images along with random 10x10 blocks of missing data before training the generator network~\cite{25}. It can be seen that as the corruption level increases, our model significantly outperforms the CE model in terms of the reconstruction quality. The architecture that we used for the CE model can be found at github \footnote[3]{https://github.com/eriklindernoren/Keras-GAN/tree/master/context\_encoder}. 
\subsection{Comparison with a Variational Autoencoder (VAE) on the task of image inpainting and denoising simultaneously}
\begin{figure}[!hbt]
    \centering
    \includegraphics[width=0.475\textwidth]{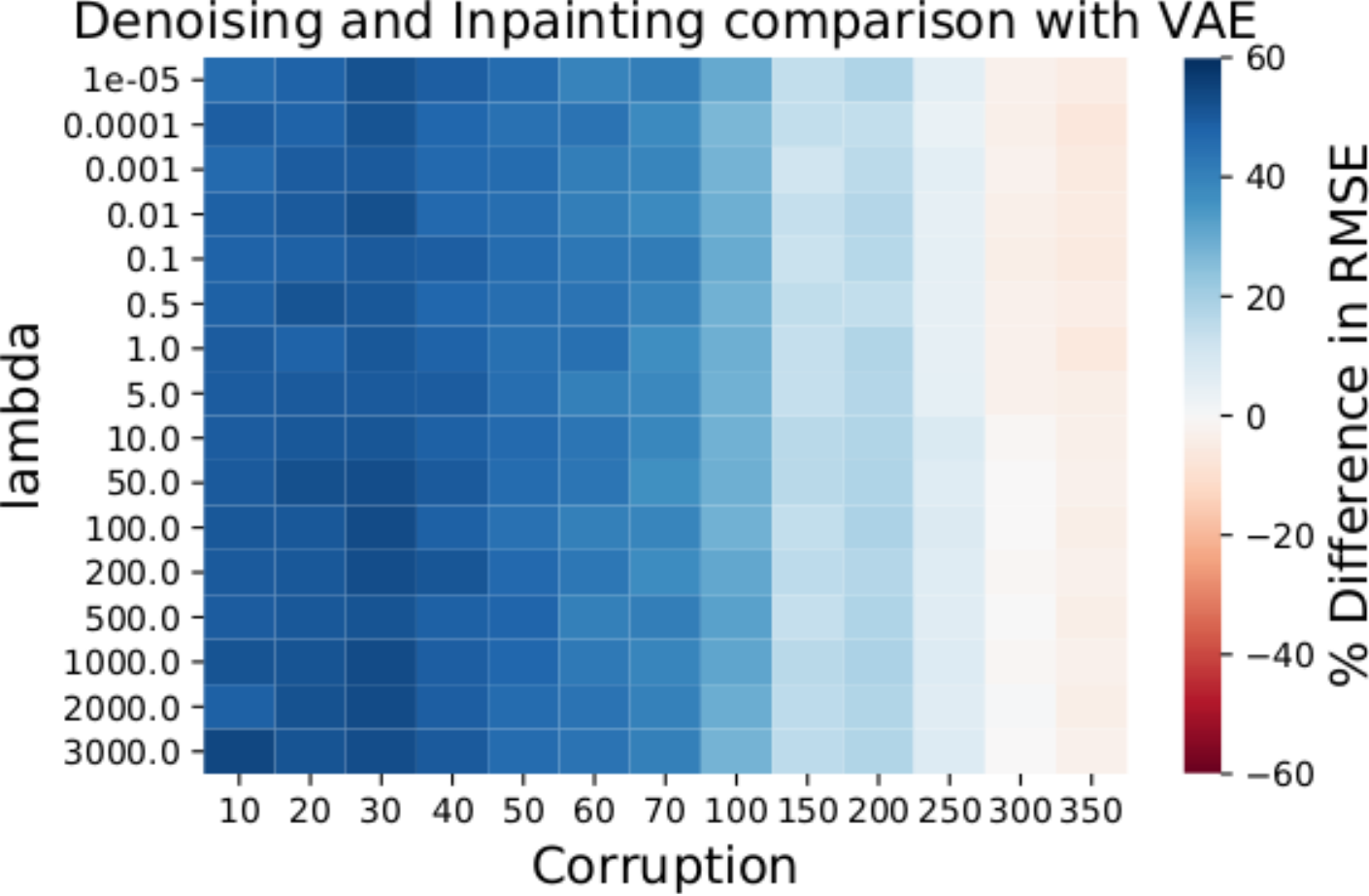}
    \caption{Comparison of our model with the reconstruction of the MNIST images in presence of noise and random 10x10 blocks of missing data using a variational autoencoder (VAE) network. The VAE model architecture and loss function is kept the same as a standard VAE model as given in ~\cite{vae}, with the modification that the input to the network is corrupted with the addition of random salt and pepper noise and missing data blocks before training. As can be seen from the colormap, our model is able to handle the reconstruction of the image, both in presence of random noise as well as missing blocks of data, as opposed to the VAE model at lower levels of corruption, while showing comparable performance at higher levels. The colormap indicates the percentage (\%) differences in RMSE between the VAE model and our model. Thus, higher the differences between the two, the better our model performs. It should be noted that the VAE model is validated based on a \emph{clean test dataset}.}
    \label{fig:plot_7}
    \vspace{-0.3cm}
\end{figure}
Variational Autoencoders (VAEs) are generative models that are modifications of the vanilla autoencoders. VAEs inherit the architecture of a standard vanilla autoencoder which allows them to sample randomly from the latent space and generate samples from the same probability distribution as the input data\cite{29}. Recent advances in the field have used the VAEs network for the task of image denoising\cite{30}. Im et al.\cite{30} use the VAE network for the task of image denoising by adding corruptions to the image before providing it as input to the network. 
Figure~\ref{fig:plot_7} shows a comparison of our model with a VAEs model on the MNIST dataset. Following the same corruption process described earlier, we corrupt the input to the VAEs model with random salt and pepper noise and also add missing 10x10 blocks of pixels at random. It can be seen that our model outperforms the VAEs model in terms of denoising and inpainting simultaneously across lower levels of corruption shown, while showing comparable performance at higher corruption levels. The corruption level here indicates the number of pixels that have been corrupted by random salt and pepper noise. It is important to note here that the VAE model input is corrupted, however, it is trained and validated using \emph{clean data}. The architecture that we used for the VAE comparison can be found at github \footnote[4]{https://github.com/lyeoni/keras-mnist-VAE}.
\subsection{Reconstruction of the noisy MNIST images using RHA}
\begin{figure*}[hbt!]
    \centering
    \includegraphics[width=0.87\textwidth]{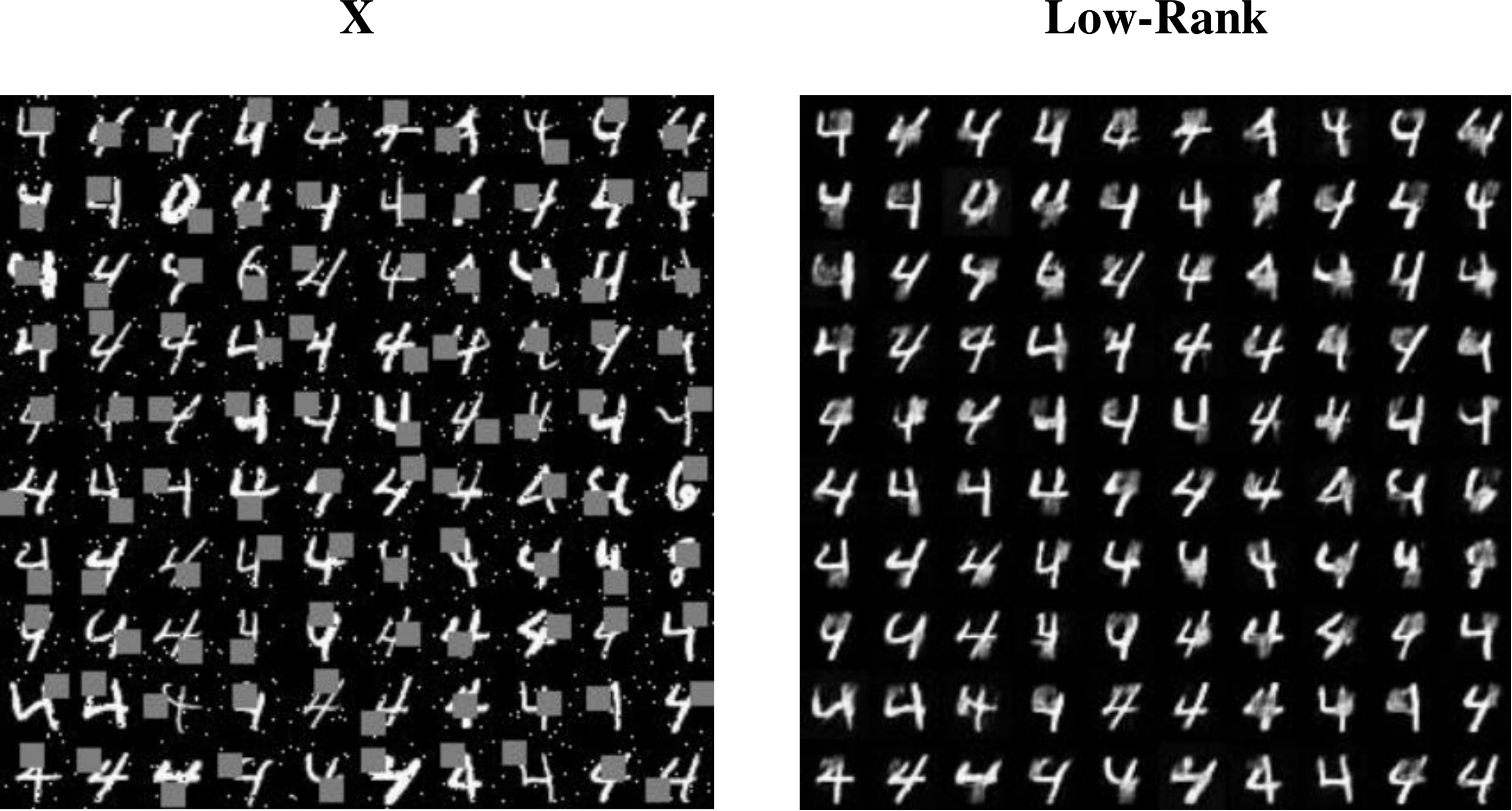}
    \caption{A sample reconstruction of the MNIST images using the RHA model. It can be seen that our model successfully inpaints and eliminates the noise simultaneously to develop a \emph{noise-free and complete} version of the image.}
    \label{fig:plot_8}
    \vspace{-0.3cm}
\end{figure*}
Figure~\ref{fig:plot_8}  shows the reconstruction of the MNIST dataset images using our RHA model, in the presence of salt and pepper noise and random 10x10 blocks of missing data. 
The image on the left is $X$, which is the corrupt input to the neural network and the image on the right is the low-rank image reconstruction that is learned by the neural network for the \emph{test dataset}. 
\subsection{Results on CIFAR10 data using RHA}
\begin{figure*}[hbt!]
    \centering
    \includegraphics[width=\textwidth]{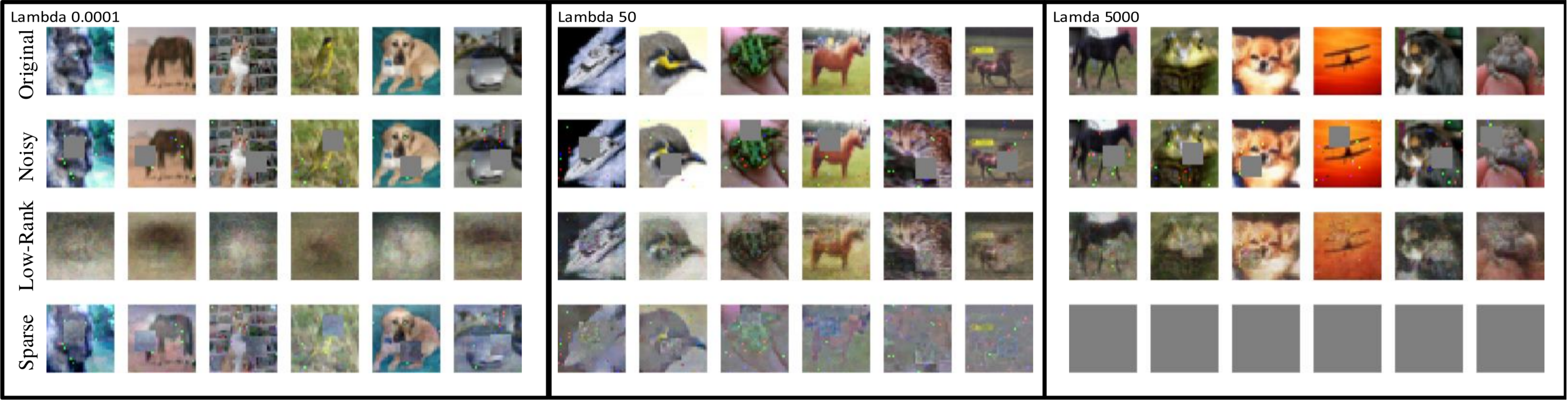}
    \caption{Results on CIFAR10 data using our RHA  model in presence of noise and random missing blocks of data of size 10x10. The first row shows the original image, the second row shows the noisy image with the masking in regions that are missing. The third and fourth rows are the low-rank and sparse reconstructions of  noisy images learnt by our model. It can be seen that depending on the values of $\lambda$, our model is able to filter out noise in the sparse matrix and generate a low-rank representation of the image, while at the same time inpaint the missing regions in the data. For lower values of $\lambda$ we impose a very small penalty on the sparse matrix and the model filters out everything in the sparse matrix. For very high values of $\lambda$, we impose a high penalty on the sparse matrix and the model fails to filter out any noise in the sparse matrix. Thus, it is important to tune $\lambda$ in a way such that the noise is filtered out in the sparse matrix, while at the same time the original \emph{noise-free and complete} data is retained in the low-rank matrix.}
    \label{fig:plot_9}
    \vspace{-0.3cm}
\end{figure*}

In Figure~\ref{fig:plot_9}, the first row shows the original image, the second row is the masked image with the addition of random noise, namely salt and pepper~\cite{25} which is fed as the input to the neural network, the third row is the low-rank matrix learnt by the neural network and the fourth row is the sparse matrix in which the noise is filtered out. The image shows the low-rank and sparse matrices learnt by the neural network for different values of $\lambda$. It can be seen that for lower values of $\lambda$ such as $0.0001$, the penalty to filter out the noise in the sparse matrix is very small and the \emph{normal} data is filtered out in the sparse matrix along with the noise. Similarly, for very high values of $\lambda$, such as $5000$, the penalty to filter out the noise and outliers in the sparse matrix is very high and the noise fails to be filtered out in the sparse matrix leading to poor reconstruction of the clean low-rank image. Hence, it is important to tune $\lambda$ and find the right value of penalty to filter out only the noise in the sparse  matrix while at the same time retaining the \emph{clean} data in the low-rank matrix . Thus, intermediate values of $\lambda$ such as $50$ are suitable and filter out the right amount of data in the sparse matrix in order to generate a low-rank matrix reconstruction that is more faithful to the original \emph{noise-free} image.
\subsection{Results on the CIFAR10 dataset using the CE model}
\begin{figure*}[ht]
    \centering
    \includegraphics[width=0.85\textwidth]{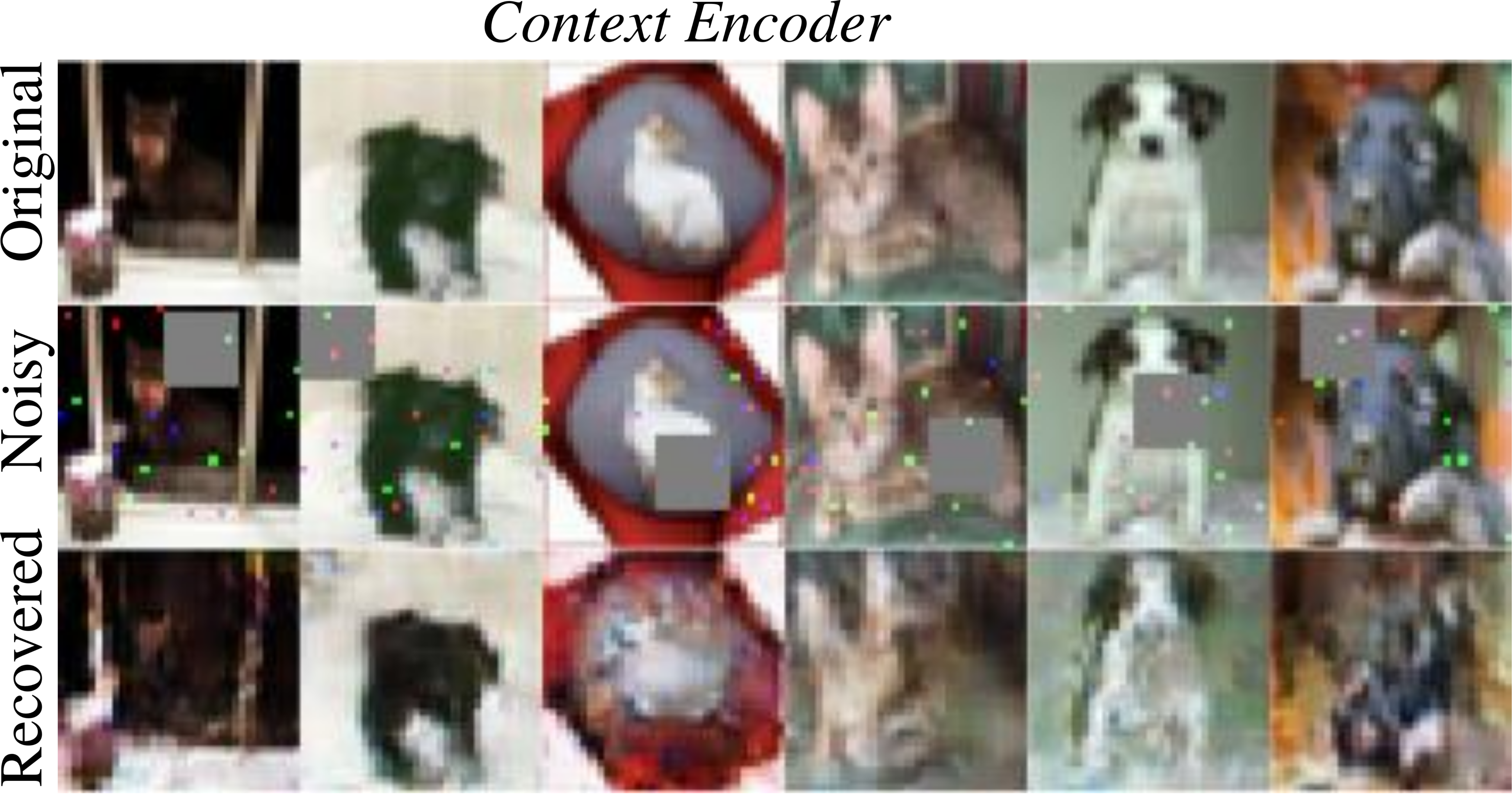}\\
    \caption{Results on CIFAR10 data using the CE model when noise is added to the input along with blocks of missing data. We add salt and pepper noise before training the model. It can be seen that the model cannot handle the noise in the input and is unable to reconstruct the \emph{noise-free} image on the task of denoising and inpainting simultaneously. The noisy image is regressed onto the ground-truth while training the generator in the GAN network.}
    \label{fig:plot_10}
    \vspace{-0.3cm}
\end{figure*}

 The poor reconstruction quality of the CIFAR10 data using the CE network is seen in Figure~\ref{fig:plot_10}. We regress the entire corrupt image onto the ground-truth version to evaluate the performance of the CE model. 
\\
\section{Conclusions}
In this paper we demonstrate a novel approach for denoising as well as image inpainting,  in which we develop a \emph{testable} version of the Robust Deep Autoencoders. We have also extended the capabilities of the model to be able to handle coherent corruptions such as random blocks of missing values in a dataset by imputing them based on the values of the surrounding pixels. This has proven useful for missing value imputation in manufacturing datasets, where we often have to deal with random blocks of missing process or chemistry data and we cannot always use naïve imputation techniques such as mean imputation~\cite{19, 20,21}. Moreover, in such scenarios, we deal with a small-sized dataset and do not want to reduce the size of the data even further by simply eliminating the rows that do not have the complete set of information. Another important application of our model is for imputing and making predictions on graph network data, for example in social networks. 
\section{Scope for future work}
We wish to extend this framework to that of anomaly detection in the future. We want to test the performance of the autoencoder network on anomaly detection in presence of  random missing blocks of data of different sizes and compare it to existing state-of-the-art approaches. In our experiments, we used the mean value to replace the pixels in the missing regions before providing the corrupt image as input to the autoencoder. In the future, we also want to experiment with different initialization values such as 0, -1, and random, to find the most suitable value to begin with before training the autoencoder for imputation and evaluate the robustness of our model to different initializations. 
In addition, the above optimization was solved using the same technique used by the RDAEs model~\cite{4} namely, a combination of backpropogation and proximal gradients~\cite{7,11,23}. We would also like to provide a theoretical justification for the failure of optimizing the model using gradient descent alone, as well as the convergence rates for the type of alternating methods we use. The $l_1$-norm optimization using backpropogation  fails to filter out the anomalies or noise, as noted in our initial experiments. This would be a direction for future work and investigation.  

\section{Acknowledgements}

We would like to thank the Advanced Casting Research Center at WPI for providing us with the funding for this project. We would also like to thank Prof. Diran Apelian for his support with this work.  

\bibliographystyle{IEEEtran}
\bibliography{references}

\begin{thebibliography}{10}
\providecommand{\url}[1]{#1}
\csname url@samestyle\endcsname
\providecommand{\newblock}{\relax}
\providecommand{\bibinfo}[2]{#2}
\providecommand{\BIBentrySTDinterwordspacing}{\spaceskip=0pt\relax}
\providecommand{\BIBentryALTinterwordstretchfactor}{4}
\providecommand{\BIBentryALTinterwordspacing}{\spaceskip=\fontdimen2\font plus
\BIBentryALTinterwordstretchfactor\fontdimen3\font minus
  \fontdimen4\font\relax}
\providecommand{\BIBforeignlanguage}[2]{{%
\expandafter\ifx\csname l@#1\endcsname\relax
\typeout{** WARNING: IEEEtran.bst: No hyphenation pattern has been}%
\typeout{** loaded for the language `#1'. Using the pattern for}%
\typeout{** the default language instead.}%
\else
\language=\csname l@#1\endcsname
\fi
#2}}
\providecommand{\BIBdecl}{\relax}
\BIBdecl

\bibitem{5}
I.~Goodfellow, Y.~Bengio, and A.~Courville, \emph{Deep learning}.\hskip 1em
  plus 0.5em minus 0.4em\relax MIT press, 2016.

\bibitem{lecun2015deep}
Y.~LeCun, Y.~Bengio, and G.~Hinton, ``Deep learning,'' \emph{nature}, vol. 521,
  no. 7553, pp. 436--444, 2015.

\bibitem{2}
P.~Vincent, H.~Larochelle, Y.~Bengio, and P.-A. Manzagol, ``Extracting and
  composing robust features with denoising autoencoders,'' in \emph{Proceedings
  of the 25th international conference on Machine learning}, 2008, pp.
  1096--1103.

\bibitem{4}
C.~Zhou and R.~C. Paffenroth, ``Anomaly detection with robust deep
  autoencoders,'' in \emph{Proceedings of the 23rd ACM SIGKDD International
  Conference on Knowledge Discovery and Data Mining}, 2017, pp. 665--674.

\bibitem{6}
E.~J. Cand{\`e}s, X.~Li, Y.~Ma, and J.~Wright, ``Robust principal component
  analysis?'' \emph{Journal of the {ACM} ({JACM})}, vol.~58, no.~3, pp. 1--37,
  2011.

\bibitem{13}
R.~C. Paffenroth, P.~C. Du~Toit, L.~L. Scharf, A.~P. Jayasumana, V.~Banadara,
  and R.~Nong, ``Space-time signal processing for distributed pattern detection
  in sensor networks,'' in \emph{Signal and Data Processing of Small Targets
  2012}, vol. 8393.\hskip 1em plus 0.5em minus 0.4em\relax International
  Society for Optics and Photonics, 2012, p. 839309.

\bibitem{28}
R.~A. Horn, ``The {H}adamard product,'' in \emph{Proc. Symp. Appl. Math},
  vol.~40, 1990, pp. 87--169.

\bibitem{12}
J.~T{\'o}th, A.~L. Nagy, and D.~Papp, ``Mathematical background,'' in
  \emph{Reaction Kinetics: Exercises, Programs and Theorems}.\hskip 1em plus
  0.5em minus 0.4em\relax Springer, 2018, pp. 359--379.

\bibitem{9}
D.~Pathak, P.~Krahenbuhl, J.~Donahue, T.~Darrell, and A.~A. Efros, ``Context
  encoders: Feature learning by inpainting,'' in \emph{Proceedings of the IEEE
  conference on computer vision and pattern recognition}, 2016, pp. 2536--2544.

\bibitem{vae}
\BIBentryALTinterwordspacing
{V}ariational autoencoders architecture. [Online]. Available:
  \url{https://github.com/lyeoni/keras-mnist-VAE}
\BIBentrySTDinterwordspacing

\bibitem{26}
\BIBentryALTinterwordspacing
O.~Lyudchik, ``{Outlier detection using autoencoders},'' Aug 2016. [Online].
  Available: \url{http://cds.cern.ch/record/2209085}
\BIBentrySTDinterwordspacing

\bibitem{27}
Y.~Ma, P.~Zhang, Y.~Cao, and L.~Guo, ``Parallel auto-encoder for efficient
  outlier detection,'' in \emph{2013 IEEE International Conference on Big
  Data}.\hskip 1em plus 0.5em minus 0.4em\relax IEEE, 2013, pp. 15--17.

\bibitem{24}
G.~L{\'o}pez, V.~Mart{\'\i}n-M{\'a}rquez, F.~Wang, and H.-K. Xu, ``Solving the
  split feasibility problem without prior knowledge of matrix norms,''
  \emph{Inverse Problems}, vol.~28, no.~8, p. 085004, 2012.

\bibitem{7}
S.~Boyd, S.~P. Boyd, and L.~Vandenberghe, \emph{Convex optimization}.\hskip 1em
  plus 0.5em minus 0.4em\relax Cambridge university press, 2004.

\bibitem{22}
B.~Efron, ``Missing data, imputation, and the bootstrap,'' \emph{Journal of the
  American Statistical Association}, vol.~89, no. 426, pp. 463--475, 1994.

\bibitem{25}
R.~H. Chan, C.-W. Ho, and M.~Nikolova, ``Salt-and-pepper noise removal by
  median-type noise detectors and detail-preserving regularization,''
  \emph{IEEE Transactions on image processing}, vol.~14, no.~10, pp.
  1479--1485, 2005.

\bibitem{17}
L.~Deng, ``The mnist database of handwritten digit images for machine learning
  research [best of the web],'' \emph{IEEE Signal Processing Magazine},
  vol.~29, no.~6, pp. 141--142, 2012.

\bibitem{18}
A.~Krizhevsky and G.~Hinton, ``Convolutional deep belief networks on
  cifar-10,'' \emph{Unpublished manuscript}, vol.~40, no.~7, pp. 1--9, 2010.

\bibitem{conae}
\BIBentryALTinterwordspacing
{C}ontext encoders architecture. [Online]. Available:
  \url{https://github.com/eriklindernoren/Keras-GAN}
\BIBentrySTDinterwordspacing

\bibitem{29}
Y.~Pu, Z.~Gan, R.~Henao, X.~Yuan, C.~Li, A.~Stevens, and L.~Carin,
  ``Variational autoencoder for deep learning of images, labels and captions,''
  in \emph{Advances in neural information processing systems}, 2016, pp.
  2352--2360.

\bibitem{30}
D.~I.~J. Im, S.~Ahn, R.~Memisevic, and Y.~Bengio, ``Denoising criterion for
  variational auto-encoding framework,'' in \emph{Thirty-First AAAI Conference
  on Artificial Intelligence}, 2017.

\bibitem{19}
L.~Monostori, ``A{I} and machine learning techniques for managing complexity,
  changes and uncertainties in manufacturing,'' \emph{Engineering applications
  of artificial intelligence}, vol.~16, no.~4, pp. 277--291, 2003.

\bibitem{20}
B.~Komer, J.~Bergstra, and C.~Eliasmith, ``Hyperopt-sklearn: automatic
  hyperparameter configuration for scikit-learn,'' in \emph{ICML workshop on
  AutoML}, vol.~9.\hskip 1em plus 0.5em minus 0.4em\relax Citeseer, 2014.

\bibitem{21}
K.~Lakshminarayan, S.~A. Harp, R.~P. Goldman, T.~Samad \emph{et~al.},
  ``Imputation of missing data using machine learning techniques.'' in
  \emph{KDD}, 1996, pp. 140--145.

\bibitem{11}
S.~Boyd, N.~Parikh, E.~Chu, B.~Peleato, J.~Eckstein \emph{et~al.},
  ``Distributed optimization and statistical learning via the alternating
  direction method of multipliers,'' \emph{Foundations and
  Trends{\textregistered} in Machine learning}, vol.~3, no.~1, pp. 1--122,
  2011.

\bibitem{23}
L.~Vinet and A.~Zhedanov, ``A ‘missing’ family of classical orthogonal
  polynomials,'' \emph{Journal of Physics A: Mathematical and Theoretical},
  vol.~44, no.~8, p. 085201, 2011.

\end{thebibliography}
\end{document}